\documentstyle[12pt]{article}
\begin{document}
\large\rm
\begin{center}
\thispagestyle{empty}
$\phantom{-}$\\
\vspace{5cm}
$\phantom{------------------}${\large\rm Preprint NPI MSU --- 96--35/442}\\
\vspace{3cm}
{\Large\rm H.F.~Goenner, G.Yu.~Bogoslovsky}\\
\vspace{2cm} 
{\LARGE\bf A CLASS OF ANISOTROPIC\\[5mm]
(FINSLER-) SPACE-TIME\\[5mm] GEOMETRIES}\\
\vspace{6.7cm}
MOSCOW 1996
\end{center}
\newpage
\pagenumbering{arabic}
\thispagestyle{empty}
\begin{center}
M.V. Lomonosov Moscow State University\\
D.V. Skobeltsyn Institute of Nuclear Physics\\
\vspace{5cm}
{\LARGE\rm A CLASS OF ANISOTROPIC\\\vspace{5mm} (FINSLER-)
SPACE-TIME GEOMETRIES}\\\vspace{1cm} {\large\rm H.F.~Goenner}\\
{\large\it Institute for Theoretical Physics, University of G\"ottingen\\
Bunsenstr.9, D-37073, G\"ottingen, Germany}\\\vspace{2mm}
{\large\rm G.Yu.~Bogoslovsky}\footnote{Supported by DFG grant No. 7/25/94
and INTAS--93--1630--EXT grant}\\
{\large\it Institute of Nuclear Physics, Moscow State University\\
119899, Moscow, Russia}\\\vspace{1cm}
Preprint NPI MSU --- 96--35/442
\end{center}
\newpage\thispagestyle{empty}
\begin{center}
{\bf Abstract}\\
\end{center}

\noindent
A particular Finsler-metric proposed in [1,2] and describing a geometry
with a preferred null direction is characterized here as belonging to 
a subclass contained in a larger class of Finsler-metrics with one or more 
preferred directions (null, space- or timelike). The metrics are classified
according to their group of isometries. These turn out to be isomorphic to
subgroups of the Poincar\'e (Lorentz-) group complemented by the generator
of a dilatation. The arising Finsler geometries may be used for the
construction of relativistic theories testing the isotropy of space. It is
shown that the Finsler space with the only preferred null direction is the
anisotropic space closest to isotropic Minkowski-space of the full class
discussed.\\

\vspace{3cm}
\noindent
H.F. Goenner e-mail: goenner@theorie.physik.uni-goettingen.de\\
G.Yu. Bogoslovsky e-mail: bogoslov@theory.npi.msu.su\\
\vspace{7.4cm}
\begin{center}
\copyright 1996, H.F. Goenner, G.Yu. Bogoslovsky
\end{center}
\newpage
\noindent {\bf 1. Introduction}

\vspace*{0.5cm}
\noindent In Refs $[1,2]$ proceeding from the group
\begin{equation}
\left.
\begin{array}{l@{\quad=\quad}ll} 
   x^{0'} &  \left(\frac{1 - v/c}{1 + v/c} \right)^{r/2} &
   \frac{x^0 - v/c~x^1}{\sqrt{1 - v^2/c^2}} \nonumber \\
   x^{1'}  & \left( \frac{1 - v/c}{1 + v/c} \right)^{r/2} &
   \frac{x^1 - v/c~x^0}{\sqrt{1 - v^2/c^2}} 
\end{array} \right\}
\end{equation}
the function 
\begin{equation}
   f = \left( \frac{x^0 - x^1}{x^0 + x^1} \right)^r \left[
   (x^0)^2 - (x^1)^2 \right]
\end{equation}
was found as a function invariant under the transformation (1)\ .
After that the corresponding 2-dimensional Finsler-metric
\[
ds = \left( \frac{d x^0 - d x^1}{d x^0 + d x^1} \right)^{r/2} \left[
(d x^0)^2 - (d x^1)^2 \right]^{1/2}
\]
was written and then generalized to 4-dimentional space 
in a coordinate independent manner with the help
of a constant vector $\nu_a$ satisfying $\nu_a \nu_b \eta^{ab} = 0:$
\begin{equation}
   ds = \left[ \frac{(\nu_a dx^a)^2}{\eta_{cd} dx^c dx^d}
\right]^{r/2} \left( \eta_{ij} dx^i dx^j \right)^{1/2}~~,
\end{equation}
where $\eta_{ab}$ is the Minkowski metric (Latin indices run form 0
to 3). Actually $\nu_a$ does {\it not}
transform as a 4-vector under the transformations belonging to an
8--parameter isometry group of the space (3); it is an {\it
absolute} preferred null vector in Finsler-space. r is called
``parameter of anisotropy"; obviously $0 \le r < 1$ must hold. In
fact, from the present tests of special relativity, $r
\stackrel{<}{\sim} 10^{-10}$. 

The purpose of this note is (i) to uniquely characterize the Finsler
metric (3) among other constructive possibilities, and (ii) to
discuss further such metrics describing geometries with inbuilt
preferred direc\-tions (not necessarily lightlike) according to the
isometry groups occurring. From such metrics theories may be
developped for experiments  testing the isotropy of space or
preferred system theories. After introducing the general class of
Finsler-metrics investigated here, in Section 2, we study the
Lie-algebra of their isometry groups. In Section 3, the metric (3)
is reobtained while in Section 4 Finsler-metrics with isometry
groups smaller than the 8-parameter group are discussed. In
Section 5 we derive the group of finite transformations for one of
the cases discussed.

\vspace*{1cm}

\noindent {\bf 2. A special class of Finsler-metrics and its isometry
group}

\vspace*{0.5cm}

\noindent We now use coordinates in which $\eta_{ab}
\stackrel{\ast}{=} {\rm diag} (+1,-1,-1,-1)$ and intro\-duce a second
symmetric, constant tensor of rank 2 $a_{cd} \left( {\rm i.e.}
~~\frac{\partial a_{cd}}{\partial x^i}~= 0 \right)$. The variable 
\begin{equation}
   v := ~~\frac{a_{cd} dx^c dx^d}{\eta_{ij} dx^i dx^j}
\end{equation}
is of degree of homogeneity zero. Any metric of the form
\begin{equation}
ds = \psi(v)~\sqrt{\eta_{ab}dx^a dx^b}
\end{equation}
with arbitrary function $\psi(v)$ such that $\lim_{v \rightarrow
\infty} ds$ exists, then is of degree one and can be used as a
Finsler metric. For the further calculations we employ the more
convenient form
\begin{equation}
   ds^2 = \phi(v) \eta_{ab} dx^a dx^b~~.
\end{equation}
$\phi = 1$ leads back to Minkowski space while $\phi = v$ leads to
the pseudo-Riemannian metric
\begin{equation}
   ds^2 = a_{cd} dx^c dx^d
\end{equation}
i.e. to a metric the rank and signature of which are not fixed from
the outset. Except for these two limiting cases, (6) represents a
proper Finsler-metric. It is tempting, however, to try and use a {\it
dual}\,\, interpretation of (6) as a pseudo-Riemannian, conformally
flat metric with conformal factor
\[
\phi \left( \frac{a_{cd} \dot{x}^c \dot{x}^d}{\eta_{ab} \dot{x}^a \dot{x}^b}
\right)~~,
\]
where $\dot{x}^c =~~\frac{dx^c}{d \lambda}~,~ \lambda$ any arbitrary 
parameter 
of the curve $x^j = x^j(\lambda)$. In this dual interpretation
$x^\lambda$ and $\dot{x}^\lambda$ would be considered as {\it
independent} variables such that $\frac{\partial \phi}{\partial xj} =
0$. From the point of view of physics, the immediate problem is
then which of the two metrics $ds^2$ and $d \ell^2 := \eta_{ab} dx^a
dx^b$ measures times and lengths. In accordance with the Finslerian
approach we take here, $ds$ is to stand for the space-time interval
measured by clocks and rigid rulers. Consequently, we will have to
cope with the occurence of a second metric (i.e. the Minkowski metric
$\eta_{ab}$) devoid of physical significance. In a way, a new type of
bi-metric theories would result with one of the metrics being
Finslerian, the other Lorentzian. However, in this paper,
only formal groundwork for physical theories to be built on is
laid. 

Starting now from metric components $\gamma_{ij}(x^k, dx^k)$ and
deriving the Lie-derivative 
\begin{equation}
   - {\cal L}_X~\gamma_{ij} := \gamma_{i' j'} (x^k, dx^k) -
\gamma_{ij} (x^k, dx^k),
\end{equation}
where $x^{'k} = x^k + \xi^k (x^\ell)$ and $X = \xi^k
~\frac{\partial}{\partial x^k}$ is the generator of the infinitesimal
coordinate transformation, we arrive at 
\begin{equation}
   - {\cal L}_X~\gamma_{ij} =  \gamma_{ij,k} \xi^k + \gamma_{i \ell} 
   \xi^\ell_{, j}~+~ \gamma_{\ell j} \xi^\ell_{,i}~+~~\frac{\partial 
   \gamma_{ij}}{\partial \dot{x}^\ell} \xi^\ell_{, k} \dot{x}^k
\end{equation}
a formula to also be found in Yano [3]. In place of $\dot{x}^\ell$ we
may also write $d x^\ell$. An isometry of the Finsler-metric with
components $\gamma_{ij}$ is defined by 
\begin{equation}
   {\cal L}_X~ \gamma_{ij} ~=~0~~.
\end{equation}
If now the Ansatz (6), i.e. $\gamma_{ij}~=~\phi(v) \eta_{ij}$ is
introduced into equation (10), we obtain
\begin{equation}
   0 = \eta_{ac} \xi^c_{,b}+ \eta_{bc} \xi^c_{,a} + 2 v~\frac{\phi
'}{\phi} \eta_{ab} \left[ \frac{a_{ij}}{a_{\ell m}\dot{x}^\ell \dot{x}^m}~~
\frac{\eta_{ij}}{\eta_{cd} \dot{x}^c \dot{x}^d} \right]~~\dot{x}^j
\xi^i_{, k} \dot{x}^k
\end{equation}
where $\phi' := ~\frac{d \phi}{d v}$. Eq. (11) may be decomposed into
\renewcommand{\theequation}{12.\arabic{equation}}
\setcounter{equation}{0}
\begin{eqnarray}\label{13}
\xi^0_{, \alpha} & - & \xi^\alpha_{, 0}  =  0~~,\\
\xi^\alpha_{, \beta} & + & \xi^\beta_{, \alpha}  = 
0~~,\\
\xi^0_{, 0} & = & \xi^1_{, 1}  =  \xi^2_{,2} = \xi^3_{,
3}~~,\\
{\rm and} ~~~ \xi^3_{,3} & + & \frac{v \phi'}{\phi}~\left[ 
\frac{a_{ij}}{a_{\ell m} \dot{x}^\ell \dot{x}^m}~-~
\frac{\eta_{ij}}{\eta_{\ell m} \dot{x}^\ell
\dot{x}^m} \right] \dot{x}^j \xi^i_{,k} \dot{x}^k = 0
\end{eqnarray}
\renewcommand{\theequation}{\arabic{equation}}
\setcounter{equation}{12}

$(\alpha, \beta = 1, 2, 3,~i, j, ... = 0,...,3)$.\\
The integration of (12.1-3) is straightforward and leads to 
\begin{eqnarray}
\xi^0 & = & \sum^3_{\beta = 1} ~f_{\beta 0} x^\beta + \alpha
(x^0),\nonumber~ \\
\xi^\gamma & = & \dot{\alpha}(x^0) x^\gamma + \sum^3_{\delta = 1}~
\Omega^\gamma{}_\delta~x^\delta + f_{\gamma 0} x^0 + f_{\gamma \gamma}
\end{eqnarray}
with $\gamma = 1, 2, 3$ and the skew matrix
\[
\Omega^\beta{}_\gamma = \left( \begin{array}{ccc}
0 & \Omega^1{}_2 & \Omega^1{}_3 \\
- \Omega^1{}_2 & 0 & \Omega^2{}_3 \\
- \Omega^1{}_3 & - \Omega^2{}_3 & 0 
                               \end{array} \right)~~.
\]
Thus, 9 constants (group prameters) $f_{\beta 0}, f_{\gamma \gamma},
\Omega^\beta{}_\gamma$ and one arbitrary function $\alpha (x^0)$
emerge. Obviously, the parameters $f_{\beta 0}$ correspond to Lorentz
boosts, $\Omega^\beta{}{\gamma}$ to space rotations while $f_{1 1},
f_{2 2}$ and $f_{3 3}$ parametrize space translations. It is to be
noted that, in principle, all parameters could be taken to also be
functions of the independent variables $\dot{x}^\ell$. The generator
connected with $\alpha (x^0)$ is 
\begin{equation}
   T_0 := \alpha (x^0)~~\frac{\partial}{\partial x^0}~+~\dot{\alpha}
(x^0) \left[ x^1 ~\frac{\partial}{\partial x^1}
~+~x^2~\frac{\partial}{\partial x^2} ~+~ x^3~\frac{\partial}{\partial
x^3} \right]~.
\end{equation}
If $T_0$ and the other nine generators are to form a Lie algebra then
\begin{equation}
   \alpha = \alpha_0 x^0 + \alpha_1
\end{equation}
with $\alpha_0, \alpha_1$ constants (or functions of $\dot{x}^\ell$)
results. Consequently, at most we obtain the 10-parameter Poincar\'e
group plus an additional dilatation generator 
\begin{equation}
\hat{T}_0 := x^0~\frac{\partial}{\partial
x^0}~+~x^1~\frac{\partial}{\partial x^1} ~+~
x^2~\frac{\partial}{\partial x^2}~+~x^3~\frac{\partial}{\partial x^3}~.
\end{equation}

Up to now eq. (12.4) has not yet been solved. In the following,
however, we shall use eq. (15) for solving eq. (12.4).

\vspace*{1cm}

\noindent {\bf 3. Most general matrix $a_{i j}$ permitting the
8-paramter isometry \phantom{aaa}  group}

\vspace*{0.5cm}

\noindent The Lie algebra generators belonging to the transformation
generalized to full 3-dimensional space, derive from [1,2]: 
\begin{eqnarray}
   \xi^0 & = & - \sum^3_{\beta = 1} \epsilon_\beta x^\beta - r 
\epsilon_3 x^0~,\nonumber \\
\xi^1 & = &  \epsilon_1 (- x^0 + x^3) + \epsilon_4 x^2 - r
\epsilon_3 x^1~,\nonumber \\
\xi^2 & = & \epsilon_2 (- x^0 + x^3) - \epsilon_4 x^1 - r
\epsilon_3 x^2~, \nonumber \\
\xi^3 & = & - \epsilon_3 x^0 - \epsilon_1 x^1 - \epsilon_2 x^2 -
r \epsilon_3 x ^3~.
\end{eqnarray}
Inspection shows that the algebra is formed by two independent
Lorentz boosts, one space rotation and a third Lorentz boost
combined with the dilatation generator $\hat{T}_0$. Insertion of eq.
(17) into eq. (12.4) leads, after a straightforward but lengthy 
integration procedure, to the following result
\[
\phi = u^{1/3} F(s, t)
\]
where $u := ~\frac{[\nu_a x^a~\nu_b \dot{x}^b]^{2 r/(1-r)}}{\eta_{c d}
\dot{x}^c \dot{x^d}}$~~, 
\begin{equation}
   s := \left( \frac{\nu_a x^a}{\nu_b \dot{x}^b} \right)^{\frac{2
r}{1-r}}~~~,~~~ t := (\nu_\ell x^\ell)^{\frac{2 r}{1-r}}~\eta_{i j}
\dot{x}^i \dot{x}^j
\end{equation}
and $\nu_a \nu_b \eta^{a b} = 0$ while $F(s, t)$ is an arbitrary
function. In order that $F$ is of degree of homogeneity zero we must
have 
\begin{equation}
   F(s, \lambda^{\frac{2}{1-r}} t) = \lambda^{(2/3)-2 r} F(s, t)~,
\end{equation}
whence follows by a well known procedure that 
\begin{equation}
\phi = \left( \frac{(\nu_a x^a)^2}{\eta_{i j} \dot{x}^i
\dot{x}^j}\right)^r~~\left( \frac{\nu_\ell \dot{x}^\ell}{\nu_j x^i}
\right)^{\frac{2 r}{3(1-r)}}~\tilde{F} \left[ \left( \frac{\nu_a
x^a}{\nu_b \dot{x}^b} \right)^{\frac{2 r}{1-r}} \right]~.
\end{equation}
$\tilde{F}(s) = s^{r - \frac{2}{3}}$ leads back to the metric of eq.
(3). It does not contain the $x^a -$ variables and thus also permits
the spacetime translation group as isometry.
Moreover, if we require this larger group to be the group of
isometries from the start, then in place of (18) the following form
of $\phi$ results
\begin{eqnarray}
   \phi & = & w^{1/2} G(z) \nonumber\\
   {\rm with} \hspace*{4cm} w & := & \frac{(\nu_a \dot{x}^a)^\frac{2
r}{1-r}}{\eta_{c d} \dot{x}^c \dot{x}^d}~~, ~~ z := (\nu_j
\dot{x}^j)^{\frac{2 r}{1-r}} \eta_{i j} \dot{x}^i \dot{x}^j~~.
\end{eqnarray}
The demand of degree of homogeneity zero for $\phi$ then uniquely
leads back to Bogoslovsky's  metric (3). It is characterized uniquely
by the requirement that one obtain a genuine Finsler-metric with the
isometry group $G_8$ described above. As can be seen from the
corresponding Lie-algebra, despite of the occurrence of the
dilatation generator $\hat{T}_0$ a subalgebra of the Poincar\'e
algebra prevails. The matrix $a_{i j} = \nu_i \nu_j$ with $\nu_a
\nu_b \eta^{a b} = 0$. 

\vspace*{1cm}

\noindent {\bf 4. Finsler-metrics with a different type of space-time
anisotropy}

\vspace*{0.5cm}

\noindent We now can approach the solution of eq. (12.4) in two ways:
either we can give the matrix $a_{i j}$ and then determine the
isometry group or, we can give the isometry group and determine $a_{i
j}$. In practice, a combination of both methods is suitable. For
example, we may start from the known subgroups of the Lorentz group
[4]. 

By inserting eqs. (13) and (15) into eq. (12.4) we obtain a set of
ten algebraic equations 
\begin{equation}
   \alpha_0 X_{i j} +~~\frac{v \phi'}{\phi}~~Y_{i j} = 0
\end{equation}
where
\begin{eqnarray*}
X_{i i} & = & a_{i i} \hspace*{1cm} (\mbox{no summation on $i$)}\\
X_{i j} & = & 2 a_{i j}~,~ i \not= j\\
\end{eqnarray*}
and
\begin{eqnarray}
Y_{0 0} & := & f_{10} a_{10} + f_{20} a_{20} + f_{30} a_{30}\nonumber\\
Y_{0 1} & := & f_{10}(a_{00} + a_{11}) + a_{21} f_{20} + a_{31} f_{30}
               - a_{20} \Omega^1{}_2  - a_{30} \Omega^1{}_3\nonumber\\
Y_{0 2} & := & f_{10} a_{12} + f_{20}(a_{00} + a_{22}) + a_{32} f_{30}
            + a_{10} \Omega^1{}_2 - a_{30} \Omega^2{}_3\nonumber\\
Y_{0 3} & := & f_{10} a_{13} + f_{20} a_{23} + f_{30}(a_{00} + a_{33}) 
              + a_{10} \Omega^1{}_3 + a_{20} \Omega^2{}_3\nonumber \\
Y_{1 1} & = & a_{01} f_{10} - a_{21} \Omega^1{}_2 - a_{31}
              \Omega^1{}_3\nonumber\\
Y_{1 2} & := & (a_{11} - a_{22}) \Omega^1{}_2 - a_{32} \Omega^1{}_3 -
              a_{31} \Omega^2{}_3 + a_{02} f_{10} + a_{01}
              f_{20}\nonumber\\
Y_{1 3} & := & a_{21} \Omega^2{}_3 + (a_{11} - a_{33}) \Omega^1{}_3 -
              a_{23} \Omega^1{}_2 + a_{03} f_{10} + a_{01} f_{30}\nonumber\\
Y_{2 3} & := & a_{12} \Omega^1{}_3 + a_{13} \Omega^1{}_2 + (a_{22} -
              a_{33}) \Omega^2{}_3 + a_{03} f_{20} + a_{02} f_{30}\nonumber\\
Y_{2 2} & := & a_{12} \Omega^1{}_2 - a_{32} \Omega^2{}_3 + a_{02}
               f_{20}\nonumber\\
Y_{3 3} & := & a_{13} \Omega^1{}_3 + a_{23} \Omega^2{}_3 + a_{03}
f_{30}~.
\end{eqnarray}
Now, if eq. (22) is to hold for arbitrary $\phi(v), \alpha_0$ must be
a function of $v$. Within the dual interpretation described in
section 2 above, this is a distinct possibility. Within our work
here, we restrict $\alpha_0$ (and all other parameters) to be
constant. Consequently, either $\alpha_0 = 0$, or $\frac{v \phi'}{\phi}
= {\rm const}$ must hold. In the first case, the dilatation generator
drops out while in the second case 
\begin{equation}
   \phi = \phi_0 v^r
\end{equation}
results with $r$ being the integration constant.

As our first example we consider the matrix $a_{ij}$ of signature
zero
\begin{equation}
   a_{ij} = \left( \begin{array}{cccc}
                   a_{00} & & 0 & \\
                          & a_{11} & & \\
                       & & a_{11} & \\
                       & 0 & & -a_{00}
                   \end{array} \right)
\end{equation}
with $a_{00} \not= a_{11}, a_{00}, a_{11} > 0$. We also set $\alpha_0 = 0$. Then $Y_{ij}
= 0$  and eq. (23) lead to the 6-parameter isometry group generated
by the space-time translations, a space-rotation $R_3$ with parameter
$\Omega^1{}_2$, and a Lorentz boost $B_3$ (parameter $f_{30})$ with
$[R_3, B_3] = 0$. Geometrically this situation corresponds to two
preferred 2-flats.

In the second example, we assume two preferred directions; i.e. we set 
\begin{equation}
   a_{ij} = 2 \nu_{(i} \mu_{j)}
\end{equation}
We work in the coordinate system in which 
\begin{eqnarray*}
   \nu_i & \stackrel{\ast}{=} & \nu_0 \delta^0_i + \delta^3_i \nu_3,
~~\mu_j~\stackrel{\ast}{=}~\mu_0 \delta^0_j + \delta^1_j \mu_1\\
a_{ij} & \stackrel{\ast}{=} & 2 \left[ \nu_0 \mu_0 \delta^0_i   
\delta^0_j  
+ \nu_0 \mu_1 \delta^0_{(i} \delta^1_{j)} + \nu_0 \mu_3 \delta^0_{(i}
\delta^3_{j)} + \nu_1 \mu_3 \delta^1_{(i} \delta^3_{j)} \right]~,
\end{eqnarray*}
where the symmetrization bracket comes with the usual factor $1/2$.
Thus, in (22) and (23) we set $\alpha_0 = 0~,~~a_{00} \cdot a_{01}
\cdot a_{03} \cdot a_{13} \not= 0$ all other $a_{ij} = 0$. Two
possible solutions emerge.
\begin{itemize}
   \item[i)] Both absolutely preferred directions $\nu_i$ and $\mu_j$
are on the null cone of the Finsler-metric (6). The isometry group is
a 2-parameter {\it abelian} subgroup of the homogeneous Lorentz group
completed with the space-time translations. The generators of the
abelian subalgebra of the Lorentz group result from the combination
of three Lorentz-boosts and three space rotations given by: 
\begin{eqnarray}
   f_{10} & = & \epsilon \Omega^1{}_3~,~~f_{20} = \epsilon \Omega^2{}_3~,
   ~~f_{30} = - \delta \Omega^1{}_3~,\nonumber\\
   \Omega^2{}_3 & = & -\epsilon \cdot \delta \Omega^1{}_2~,
\end{eqnarray}
with $\epsilon^2 = \delta^2 = 1~,~~\nu_0 = \epsilon \nu_3~,~\mu_0 =
\delta \mu_1$~.

$\Omega^1{}_2$ and $\Omega^1{}_3$ are the independent parameters.
\item[ii)] None of the two preferred absolute directions lies on the
light cone. The isometry group is a one-parameter subgroup of the
homogeneous Lorentz group plus the space-time translation group. The
single generator of the first mentioned subgroup  is a combina\-tion of
a Lorentz boost and two rotations
\begin{eqnarray}
   f_{20} & = & -~\frac{\mu_1}{\mu_0}~\Omega^1{}_2~~,~~~f_{10} = f_{30}
                  = 0\nonumber\\
   \Omega^2{}_3 & = & -~\frac{\nu_0}{\nu_3}~~\frac{\mu_1}{\mu_0}~
   \Omega^1{}_2~,~~\Omega^1{}_3 = 0~.
\end{eqnarray}

As a final example, we consider a preferred {\it spacelike}
direction $\nu_a$. Using coordinates in which $\nu_a
\stackrel{\ast}{=} \delta^1_a$, i.e. $\eta^{ab} \nu_a \nu_b = -
1$, we obtain from $Y_{ij} = 0$ with $a_{11} = 1$ and all other
$a_{ij} = 0$: 
\[
f_{10} = 0~~,~\Omega^1{}_2 = \Omega^1{}_3 = 0~.
\]
The 3-parameter subgroup with parameters $f_{20}, f_{30}$ and
$\Omega^2{}_3$ \\ 
remains. In contrast, the preferred {\it timelike}
direction $\nu_a \stackrel{\ast}{=} \delta^0_a$ would retain the
group of space-rotations as an isometry group but loose all boosts.

\vspace*{1cm}

\noindent {\bf 5. Finite transformations}

\vspace*{0.5cm}

It is no problem to find the groups of finite transformations
belong\-ing to (27) or (28). In the first case, after taking $\epsilon
= \delta = \mu_1 = v_3 = 1$, the generators of the abelian subgroup of
the Lorentz group are
\begin{equation}
   X_1 = (x^1 - x^3)~\frac{\partial}{\partial x^0}~+ (x^0 +
x^3)~\frac{\partial}{\partial x^1}~- (x^0 +
x^1)~\frac{\partial}{\partial x^3}
\end{equation}
and
\begin{equation}
   X_2 = x^2 \left( \frac{\partial}{\partial x^0}~-~\frac{\partial}{\partial 
   x^1}~-~\frac{\partial}{\partial x^3} \right)~+~(x^0 + x^1 + 
   x^3)~\frac{\partial}{\partial x^2}~.
\end{equation}
By summing up the Taylor series
\[
x^i + u~X_\alpha x^i + ~\frac{u^2}{2!} (X_\alpha)^2 x^i + ...\quad
(\alpha = 1,2)
\]
we obtain the {\em finite} transformations corresponding to eqs. (29)
and (30), respectively
\begin{eqnarray}
  & x^{\alpha '}  =  x^\alpha + \sinh u [\delta^\alpha_0 (x^1 - x^3) +
\delta^\alpha_1  (x^0 + x^3) - \delta^\alpha_3 (x^0 + x^1)] & \nonumber
\\ 
&  +   (\cosh u - 1) [\delta^\alpha_0 (x^1 + x^3 + 2 x^0) - 
\delta^\alpha_1 (x^0 + x^3) - \delta^\alpha_3 (x^0 + x^1)] &
\end{eqnarray}
and 
\begin{eqnarray}
   x^{\alpha '} & = & x^\alpha + \sin \varphi [x^2 (\delta^\alpha_0 -
\delta^\alpha_1 - \delta^\alpha_3) + (x^0 + x^1 + x^3)
\delta^\alpha_2]\nonumber \\ 
& & + (1 - \cos \varphi) [x^0 + x^1 + x^3)(\delta^\alpha_0 - 
\delta^\alpha_1 - \delta^\alpha_3) - x^2 \delta^\alpha_2].
\end{eqnarray}
$u$ and $\varphi$ are the group parameters. We set $\sinh u = \gamma
\upsilon/c, \cosh u = \gamma$ with $\gamma := (1 - \upsilon^2/c^2)^{-1/2}$ 
and introduce new coordinates
\[
y^0 = x^0 + x^1 , y^1 = x^1 , y^2 = x^2 , y^3 = x^0 + x^3.
\]
Eq. (31) then can be written as
\begin{eqnarray}
   y^{0'} & = & \gamma(1 +~\frac{v}{c}) y^0 \nonumber \\
   y^{1'} & = & y^1 \left( 1 - \sqrt{\frac{1 - \upsilon/c}{1 +
                \upsilon/c}} \right) y^3 \nonumber \\
   y^{2'} & = & y^2 \nonumber \\
   y^{3'} & = & \gamma \left( 1 -~\frac{v}{c} \right) y^3. 
\end{eqnarray}
Similarly, by introducing new coordinates 
\[
y^0 = x^0 + x^1~,~y^1 = x^0 + x^1 +x^3~,~y^2 = -x^2~,~y^3 = x^0 + x^3
\]
we obtain from eq. (32)
\begin{eqnarray}
   y^{0'} & = & y^0 \nonumber\\
   y^{2'} & = & \cos \varphi y^1 + \sin \varphi y^2 \nonumber\\
   y^{2'} & = & - \sin \varphi y^1 + \cos \varphi y^2 \nonumber\\
   y^{3'} & = & y^3~.
\end{eqnarray}
The space-time translations have to be added to eqs. (33) and (34).
An invariant function thus is 
\[
f \left( \frac{y^0 y^3}{(y^0)^2 - (y^1)^2 - (y^2)^2 - (y^3)^2}
\right) [(y^0)^2 - (y^1)^2 - (y^2)^2 - (y^3)^2]
\]
such that we arrive at Finsler metric
\[
ds = \left[ \frac{d y_0 d y_3}{\eta_{ab} d y^a d y^b} \right]^{s/2}
\sqrt{\eta_{ij} d y^i d y^j}
\]
with the anisotropy parameter $s$.

\vspace*{0.5cm}

\noindent {\bf 6. Conclusions}

\noindent By introducing space-time geometries with preferred
directions, in general, Lorentz boosts or space rotations are lost as
space-time isometries. Consequently, angular momentum conservation
and/or conservation of the center of mass is no longer fully
guaranteed. Thus, such geometries may be used to construct
relativistic theories with anisotropic space or with preferred
reference systems (preferred directions in velocity space). To be
close enough to present empirical data, Finsler-geometries with the
largest possible isometry groups (as subgroups of the Poincar\'e
group) are to be preferred. As no 5-parameter subgroup of the
homogeneous Lorentz-transformations exists [4], the largest
nontrivial such isometry group is an 8-parameter group.
The Finslerian special relativity theory of Refs $[1,2]$ is built on such
an 8-parameter group. It is unique, because only one non-isomorphic
4-parameter subgroup of the homogeneous Lorentz group does exist.
There are, however, other possibilities for Finsler geometries with
7-parameter isometry groups as exemplified by the formalism of this
paper and which could also be used for the construction of relativistic 
test theories for anisotropy or a preferred system of reference. 
\end{itemize}
{\bf References}

\begin{itemize}
   \item[[1]] G.Yu. Bogoslovsky, {\it Theory of Locally Anisotropic
Space-Time}, Moscow State University Press, Moscow (1992).
 \item[[2]] G.Yu. Bogoslovsky, Fortschr. Phys. {\bf 42} (1994)
143-193:  {\it A Viable Model of Locally Anisotropic Space-Time and the
Finslerian Generalization of the Relativity Theory}\ ; Phys. Part. Nucl.
{\bf 24} (1993) 354-378:  {\it Finsler Model of Space-Time}.
 \item[[3]] K. Yano, {\it The Theory of Lie Derivatives and its
Applications}, North Holland, Amsterdam (1955).
 \item[[4]] P. Winternitz and I. Fri\v{s}, Sov. J. Nucl. Phys. {\bf 1}
(1965) 636-641:  {\it Invariant Expansions of
Relativistic Amplitudes and Subgroups of the Proper Lorentz Group}.
\end{itemize}
\end{document}